\documentclass[journal=nalefd,manuscript=letter]{achemso}
\usepackage[version=3]{mhchem}
\usepackage{amsmath}

\author{Mason J. Guffey}
\affiliation{Alpha Ring Intl. Lmtd.}
\email{mason@alpharing.com}
\author{Alfred Y. Wong}


\title[]{Ponderomotive Screening of Nuclear Fusion Reactions Based on Localized Surface Plasmon Resonance}
\makeatletter
\ifxetex
  \usepackage[setpagesize=false, 
              unicode=false, 
              xetex]{hyperref}
\else
  \usepackage[unicode=true]{hyperref}
\fi
\hypersetup{breaklinks=true,
            bookmarks=true,
            pdfauthor={},
            pdftitle={},
            colorlinks=true,
            urlcolor=blue,
            linkcolor=magenta,
            pdfborder={0 0 0}}
\begin{document}
\begin{abstract}
A scheme is presented to catalyze nuclear fusion by the excitation of the localized surface plasmon resonance (LSPR) of an Au nanobipyramid (NBP) by ultrafast (femtosecond) laser pulses. The effect utilizes the exceptionally high electric field enhancement provided by the Au NBP LSPR, localized to nm\textsuperscript{3} volumes and fs time scales, to produce a ponderomotive force that in turn provides for screening energies near the Au NBP tip. Discrete-dipole approximation (DDA) simulations are presented to support the proposed scheme. Calculations made with conservative parameters suggest that the effect should be observable in a laboratory setting using commercially available ultrafast lasers.
\end{abstract}
\begin{tocentry}
\includegraphics{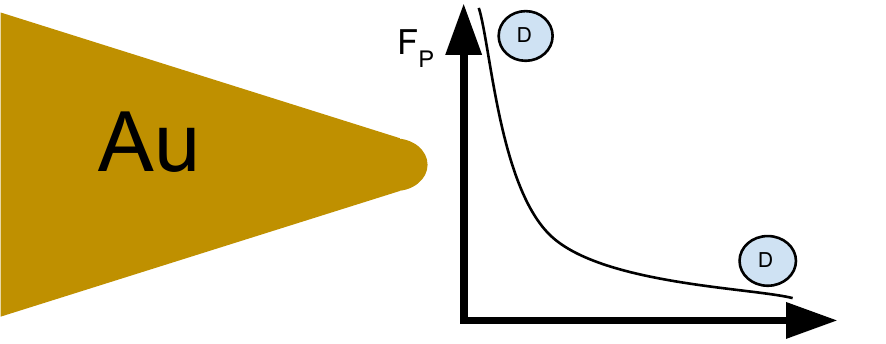}
\end{tocentry}

\hypertarget{introduction}{%
\section{Introduction}\label{introduction}}

The field of plasmonics is concerned with the manipulation and confinement of electromagnetic energy in nano-scale geometries.\textsuperscript{1} This is generally accomplished with systems consisting of Au or Ag nanomaterials optically excited by surface plasmon resonance (SPR). In the case of thin metal films, this excitation takes the form of electron density waves propagating along the metal-dielectric interface. In the case of metal nanoparticles, size, shape, and local dielectric environment determine the conditions for the so-called localized surface plasmon resonance (LSPR), in which conduction electrons oscillate coherently with the driving field inside the volume of the nanoparticle.

In general for plasmon surface resonance there is significant enhancement of the local electric field at the metal-dielectric interface. For an incident electromagnetic wave with field E\textsubscript{0}, the local electric field at the metal surface on resonance E\textsubscript{SP} can be enhanced by factors E\textsubscript{SP}/E\textsubscript{0} of anywhere from 5-10 for Au thin films,\textsuperscript{2} to factors of 200 or more for sharp-tipped nanostructures such as Au nano-bipyramids (NBP's)\textsuperscript{3}.

The fundamental physical cause for the observed electric field enhancement is the bunching of conduction electrons at the metal surface driven by the incident field. The electrons accumulate at sharp dielectric discontinuities, such as the surface of the film in the case of SPR or the edges or tips of nanoparticles in LSPR, at every cycle in phase with the driving optical field. This coherent oscillation of the conduction electrons in the metal may persist for some 10's of femtoseconds after the driving field has ceased, followed by thermalization to a so-called ``hot electron'' distribution\textsuperscript{4}.

When combined with ultrafast (femtosecond pulsed) laser systems delivering high peak powers, plasmonic nanomaterials can exhibit surface electric fields of 10\textsuperscript{9} V/m or higher. These large fields can give rise to a number of nonlinear optical effects that have been observed, including second harmonic generation (SHG),\textsuperscript{5} ponderomotive acceleration of electrons,\textsuperscript{6} and other second- and higher-order nonlinearities.

In particular, evidence for significant ponderomotive forces produced by plasmon resonances have been shown in several previous reports. Irvine and Elezabbi demonstrated that the excitation of a thin Ag film with an ultrafast laser pulse will allow electrons to tunnel through the metal surface and become accelerated to keV energies via the ponderomotive force.\textsuperscript{6,7} A similar report by Dombi et. al.~shows the emission and acceleration of electrons from lithographically-fabricated Au nanoparticles.\textsuperscript{8}

The unique aspect of surface plasmon resonance phenomena to concentrate and manipulate electrons in nanoscale volumes has found applications in areas such as biosensing, cancer therapy, and solar energy conversion. However the application of SPR and LSPR phenomena to nuclear science has remained largely unexplored. Multiple experimental studies of beam-target nuclear fusion have established the ability of bound electrons in various metallic systems to provide an effective screening of up to \textasciitilde800 eV for nuclear fusion reactions.\textsuperscript{9,10} Given these reports of screening from equilibrium electron densities, it is reasonable to expect that dynamically enhanced non-equilibrium electron densities arising from LSPR excitation might further enhance the screening of nuclear fusion reactions.\textsuperscript{11,12}

Here we predict based on electrodynamic simulations that the ponderomotive forces present at the tips of Au NBPs excited by ultrashort laser pulses are capable of providing multi-keV screening energies for light-element nuclear fusion, equivalent to or greater than that of muon-catalyzed fusion at 5.6 keV.

\hypertarget{results-and-discussion}{%
\section{Results and Discussion}\label{results-and-discussion}}

\begin{figure}

{\centering \includegraphics[width=0.4\linewidth]{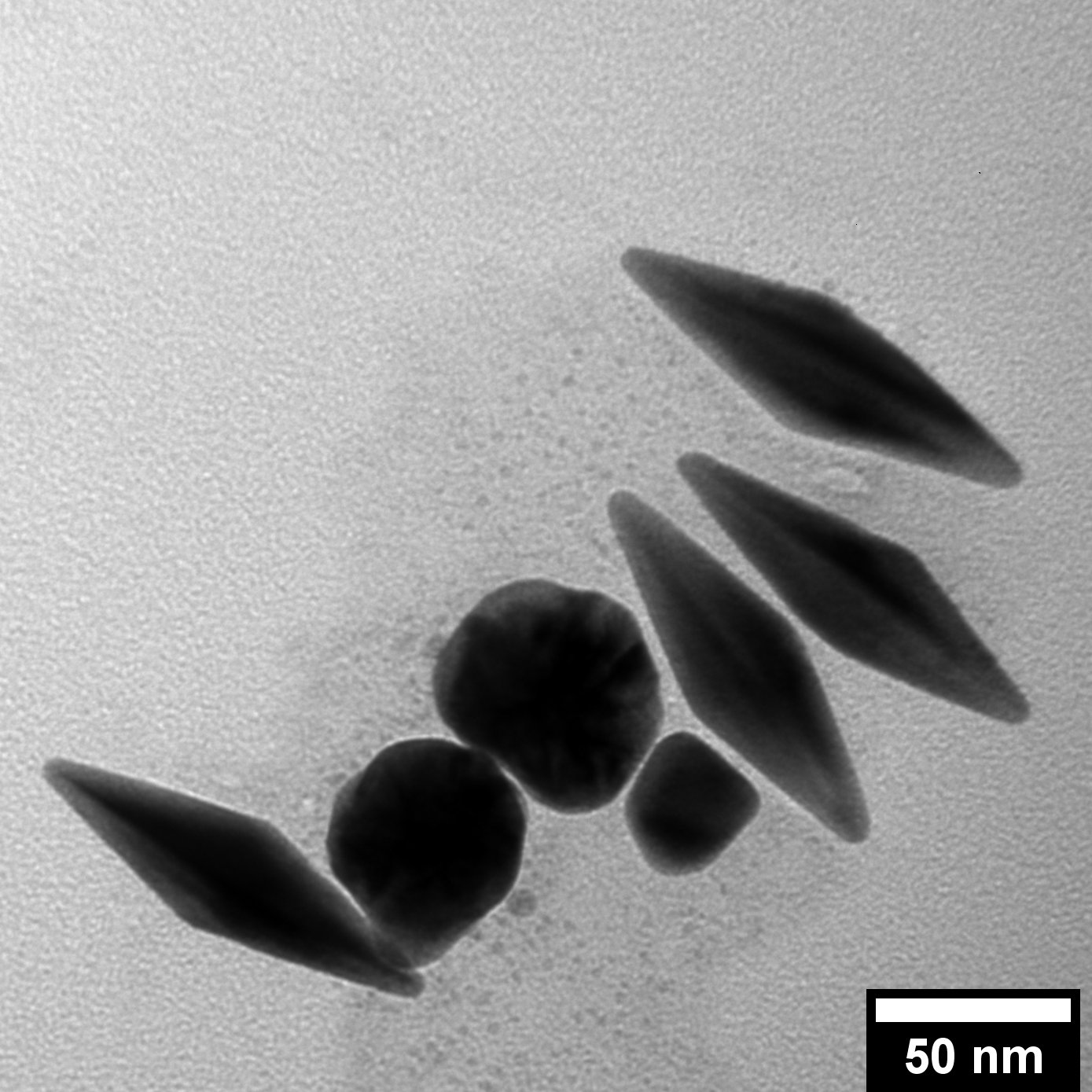} 

}

\caption{Transmission electron micrograph of Au NBP's along with spheroidal particles}\label{fig:AuBipyTEM}
\end{figure}

The Au NBP (Figure \ref{fig:AuBipyTEM}) is particularly well suited for the purpose of generating large non-equilibrium electric fields in nanoscale volumes. Au NBP's are synthesized using aqueous colloidal chemistry with cetyltrimethylammonium bromide (CTAB) as a ligand. The NBP structure consists of two pentagonal pyramids joined at the base.\textsuperscript{13} Due to the symmetric, tapered tip structure, the main dipolar longitudinal LSPR mode of the Au NBP concentrates conduction electrons at the tips during each half-cycle of the driving field.

As is the case with Au nanorods, the greater-than-unity aspect ratio of the Au NBP's leads to a spectral red-shift in the longitudinal LSPR mode, resulting in an extinction peak in the near-IR.\textsuperscript{14}

The optical response of the Au NBP is accurately simulated using the discrete dipole approximation (DDA) method, a frequency-domain technique that discretizes an arbitrary spatial geometry into a 3-dimensional array of coupled dipoles, and then determines the optical absorption and scattering properties of the entire coupled system.\textsuperscript{15} Several codes exist for running DDA simulations, here we used the open-source nanoDDSCAT resource freely available at nanoHub.org.\textsuperscript{16}

\begin{figure}

{\centering \includegraphics[width=0.65\linewidth]{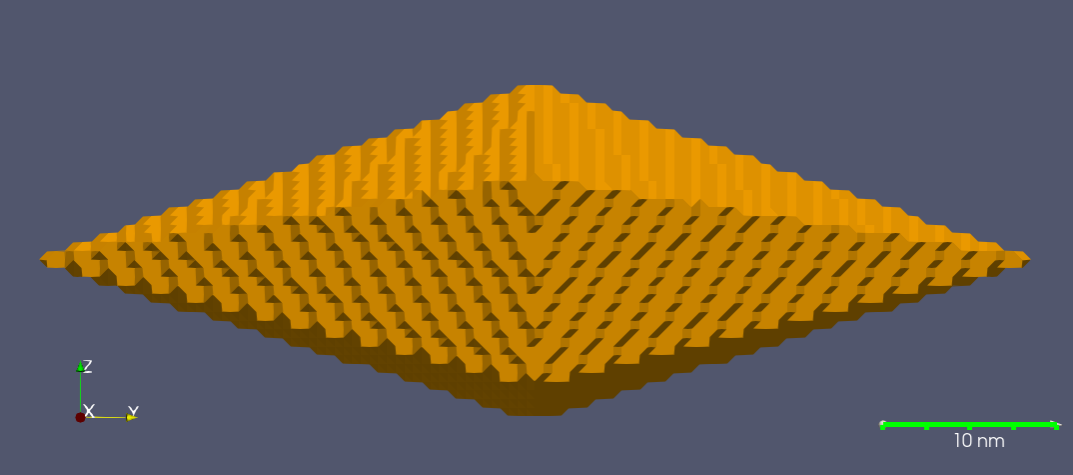} \includegraphics[width=0.3\linewidth]{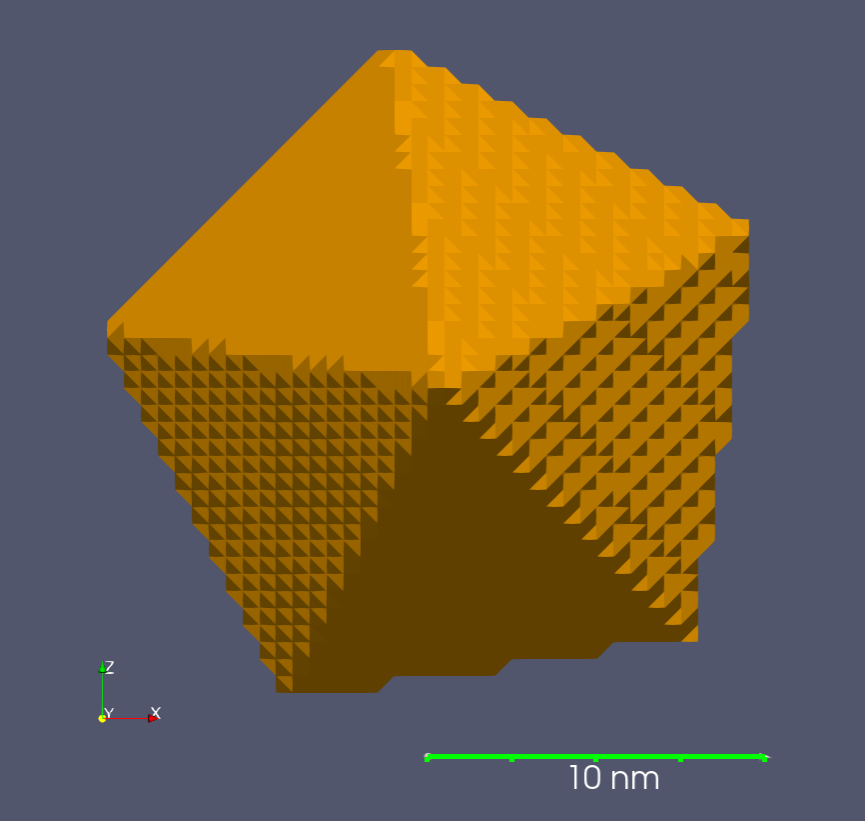} 

}

\caption{Bipyramid structural model used in DDA simulations.}\label{fig:DDAGeomFigures}
\end{figure}

The simulation was performed as follows. First, an approximate 3D structural model of a bipyramid was created in the Wavefront OBJ file format using Autodesk® Tinkercad™. An aspect ratio of \textasciitilde{} 3 was created by merging two pentagonal pyramids with a height of 30 nm and a width of 10 nm at the base. The tips were truncated by 1 nm to more closely match typical NBP geometries derived by electron microscopy. This OBJ file was then converted into a format compatible with the nanoDDSCAT codebase by using the DDSCAT Convert tool also available on nanoHub.org, see Figure \ref{fig:DDAGeomFigures}.\textsuperscript{17} A resolution of 0.5 nm was employed for the structural model, which achieved good agreement in terms of both spectral response and E-field enhancement with previous results.\textsuperscript{3}

One output of the DDA simulation is the spatially-resolved local electric field enhancement at resonance, or \textbar E/E\textsubscript{0}\textbar. This is the factor by which the electric field near the surface of the Au NBP is enhanced over the incident electric field E\textsubscript{0} due to resonant excitation of the metal conduction electrons.

\begin{figure}

{\centering \includegraphics[width=0.54\linewidth]{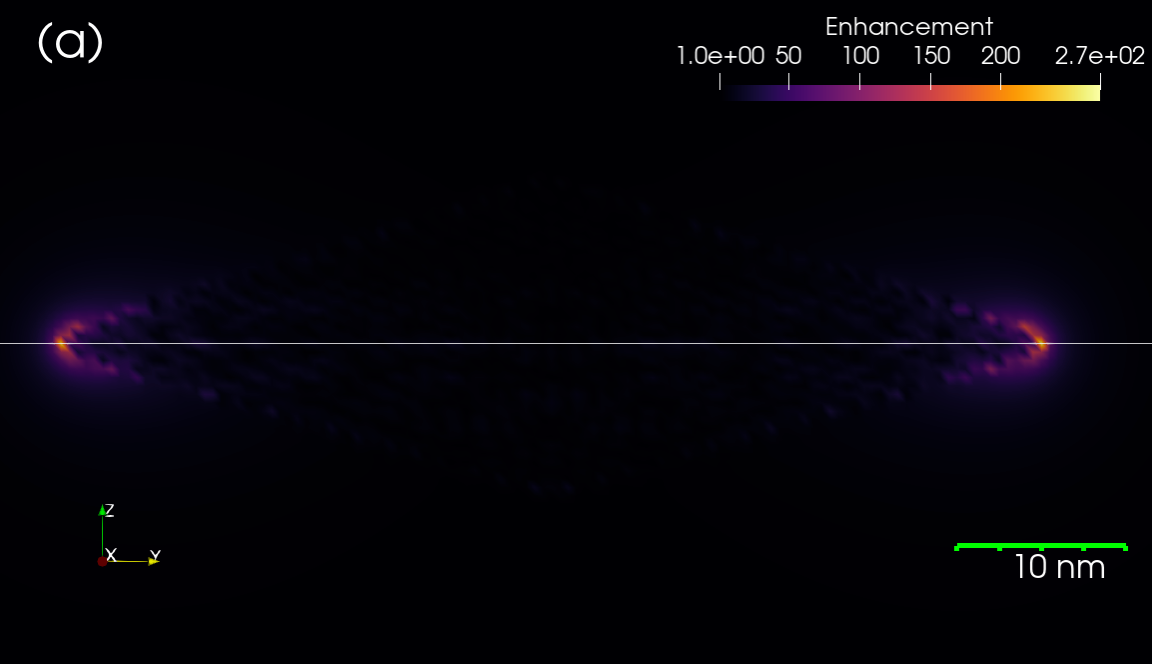} \includegraphics[width=0.45\linewidth]{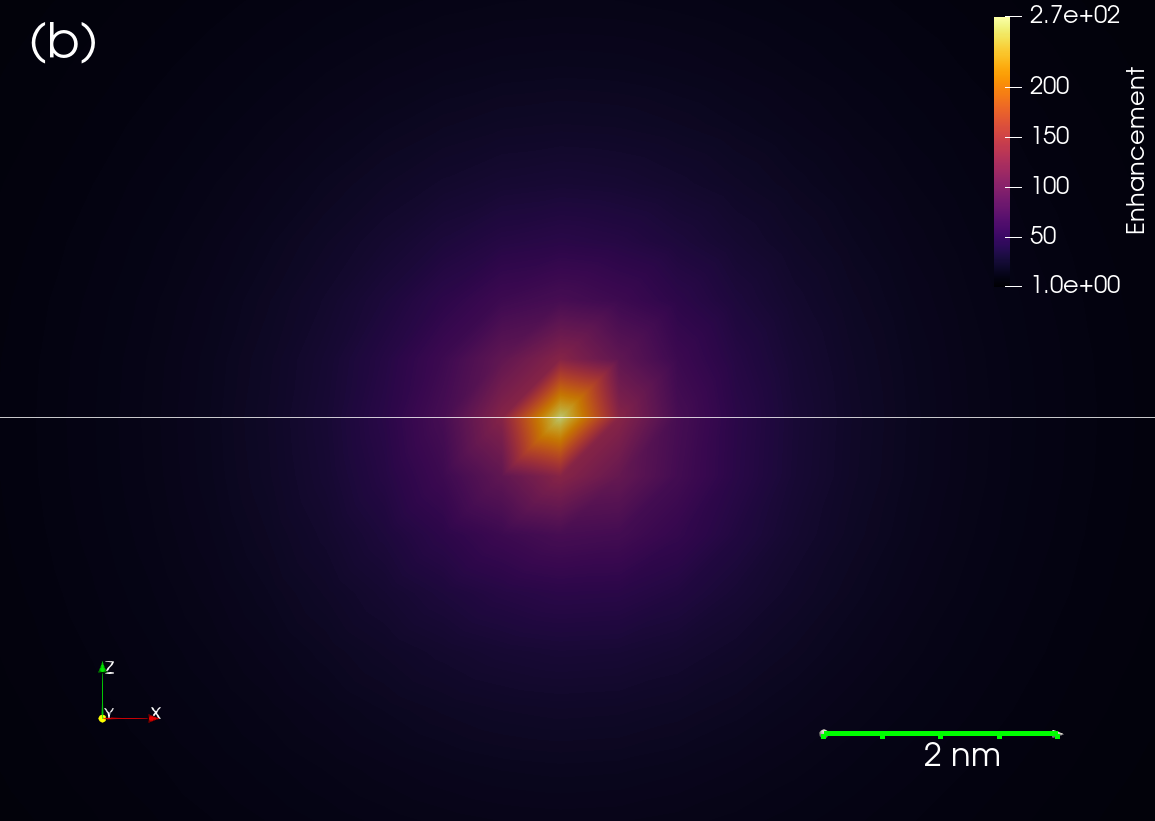} 

}

\caption{False color maps of E/E\textsubscript{0} for the (a) longitudinal plane of the Au NBP and (b) a plane normal to the long axis of the NBP passing through the tip. White lines correspond to field profiles for the next figure.}\label{fig:EFieldMaps}
\end{figure}

\begin{figure}

{\centering \includegraphics[width=0.49\linewidth]{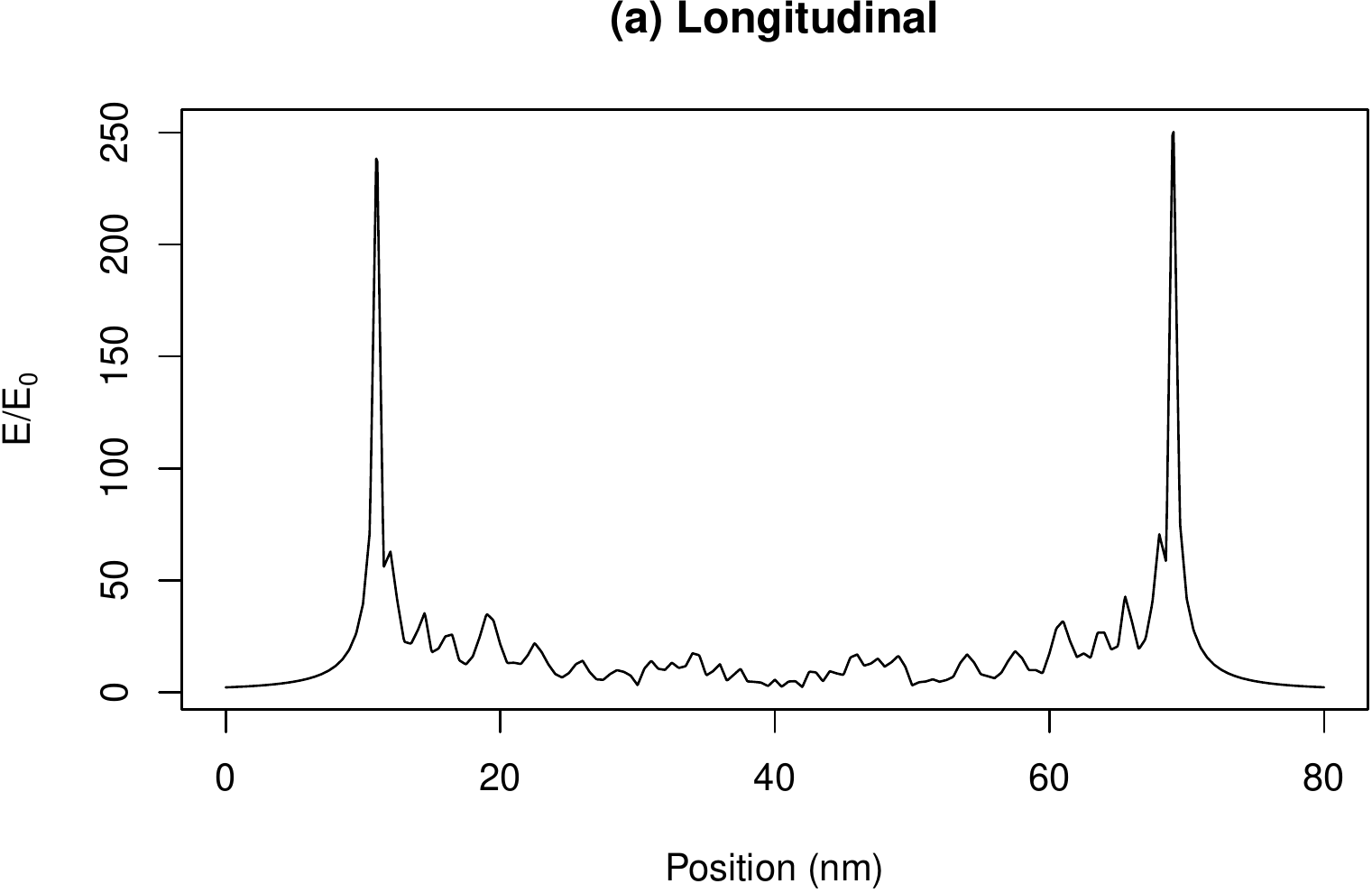} \includegraphics[width=0.49\linewidth]{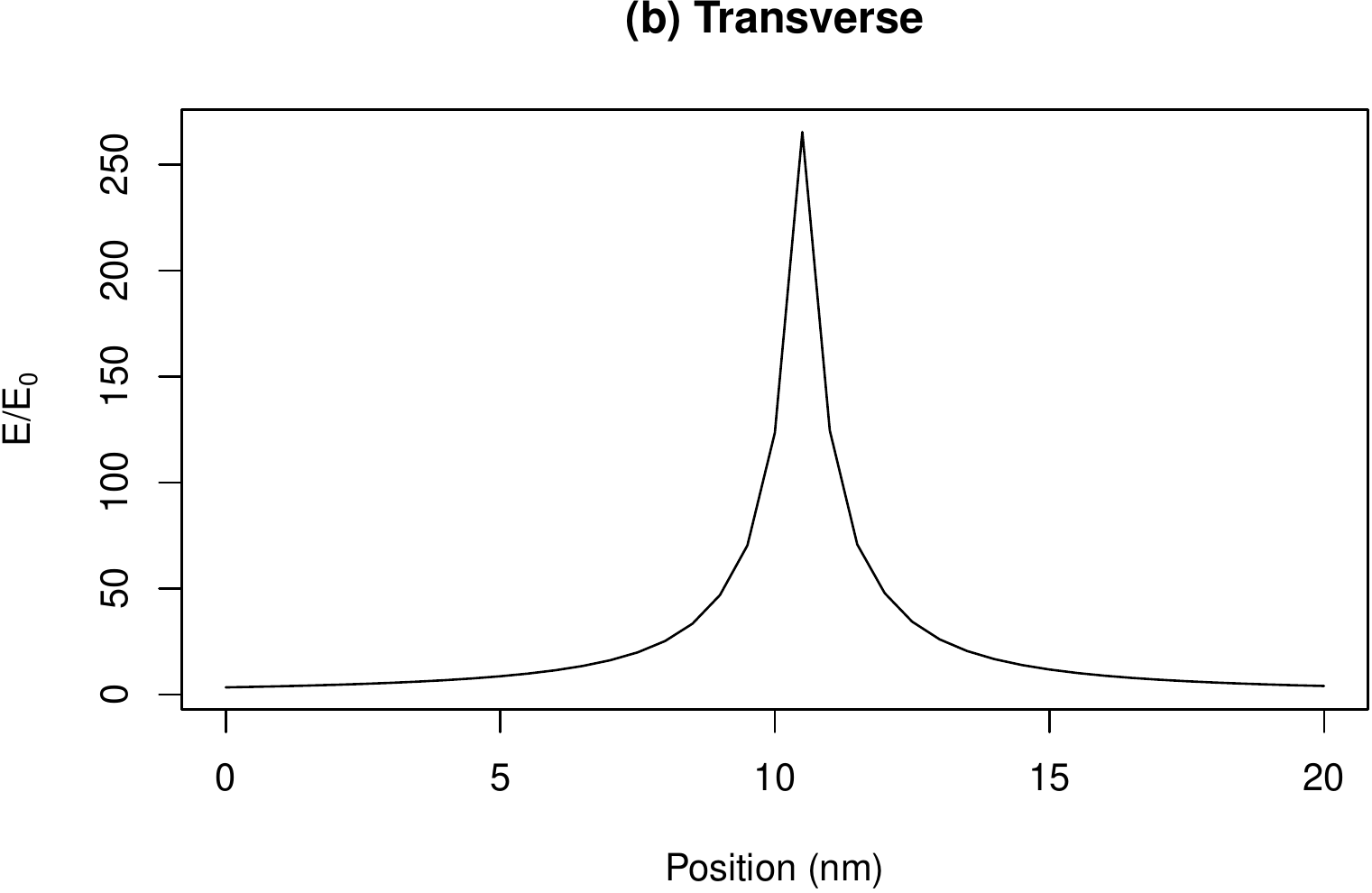} 

}

\caption{Linear profiles of E/E\textsubscript{0} for the (a) longitudinal axis of the Au NBP and (b) a line perpendicular to the long axis of the NBP and passing through the tip, as shown in the previous figure.}\label{fig:ImportPlotEData}
\end{figure}

Figure \ref{fig:EFieldMaps} shows false-color mappings of the electric field enhancement for a plane containing the long axis of the NBP as well as a plane perpendicular to the NBP long axis containing the high-field region near the tip. Figure \ref{fig:ImportPlotEData} shows linear profiles of the enhancement factor through along the white lines shown in Figure \ref{fig:EFieldMaps}, highlighting the maximum electric field enhancement of \textasciitilde{} 250.

The ponderomotive force is a repulsive force acting on a charged particle within an electric field that is both oscillating and spatially inhomogeneous. Specifically, the ponderomotive force acting on electrons is proportional to the gradient of the electric field squared, as in

\begin{equation}
\mathbf{F}_{p} = -\frac{e^2}{4m_e\omega^2}\nabla(E^2)
\label{eq:FpNabla}
\end{equation}

where \(m_e\), \(\omega\) and \emph{E} are the electron mass, optical frequency and peak electric field magnitude respectively. At a position near the tip of the Au NBP, there are two directions that have non-zero spatial derivatives of \emph{E}, these being the Y and X directions in the axes shown in Figures \ref{fig:DDAGeomFigures} and \ref{fig:EFieldMaps}. Thus \textbf{F}\textsubscript{p} is given by

\begin{equation}
\mathbf{F}_{p} = -\frac{e^2}{4m_e\omega^2}[\frac{d(E^2)}{dy}+\frac{d(E^2)}{dx}]
\label{eq:FpCartesian}
\end{equation}

We can use the E-field derived from the DDA calculation to calculate the ponderomotive force based on Equation \eqref{eq:FpCartesian}.

To do this, we first need to define characteristics of an incident optical excitation, i.e.~a femtosecond laser pulse. Consider a commercially available Spectra-Physics Mai Tai® laser system producing laser pulses at 850 nm center wavelength with 70 fs peak width and 10 nanojoules of energy per pulse. This results in a peak laser power \ensuremath{1.43\times 10^{5}} W. Focusing this beam into a spot size diameter of 5 um, the peak incident intensity I\textsubscript{0} is \ensuremath{7.28\times 10^{11}} W/cm\textsuperscript{2}. While a CW (or even nanosecond-pulsed) laser with this power density would most likely melt metal nanostructures, previous reports have shown a femtosecond-pulsed source with a comparable peak intensity is at or below the damage threshold.\textsuperscript{5} Using the relationship between optical intensity and electric field,

\begin{equation}
I_0= \frac{1}{2} c \varepsilon_0 n  E_0^2  \text{,}
\label{eq:intensity}
\end{equation}

the incident electric field E\textsubscript{0} can be calculated as \ensuremath{2.03\times 10^{9}} V/m.

Using the electric field enhancement values derived from the DDA simulation, the laser parameters from the previous section and Equation \eqref{eq:FpCartesian}, the \emph{y} and \emph{x} components of the ponderomotive force \textbf{F}\textsubscript{py} and \textbf{F}\textsubscript{px} at the tip surface can be calculated as \ensuremath{1.05\times 10^{-6}} and \ensuremath{2.14\times 10^{-7}} N respectively. As expected the \emph{y} component of the ponderomotive force is greater than the \emph{x} because this is the direction parallel to the long axis of the Au NBP. Assuming that this force acts over a distance of 1 nm, the magnitude of the screening energy \emph{E}\textsubscript{s} is 7.9 keV. It is noted that the force on electrons will be also transmitted to nearby nuclei, thus providing for a screening potential. The screening energy calculated here is comparable to that for muon-catalyzed fusion.

The reaction rate will now be estimated based on a simplified physical picture. Let us assume the Au NBP surface is coated with a deuterium-enriched alkanethiol molecule (ligand) such as dodecanethiol-d25 (C\textsubscript{12}HD\textsubscript{25}S).\textsuperscript{18} The transient screening potential established in the previous section will repel electrons from the high-field zone near the particle tip, a volume consisting primarily of the ligand layer. Due to the near-solid density of this environment, the electrons will not be able to travel very far from the tip. Subsequently, residual field-ionized deuterium atoms in the ligand structure will move towards the non-equilibrium electron density and collide with other deuterium atoms. All of the aforementioned steps would take place within the femtosecond pulse envelope, which would allow for sufficient time for both electron and ion (D\textsuperscript{+}) motion.

Assuming that there are no ``recoil-like'' re-collisions that take place, or in other words that all affected deuterium ions get one shot to fuse, we can estimate the fusion reaction probability \emph{P} for a single laser pulse as

\begin{equation}
P = N_D \sigma_{DD} n_D x
\label{eq:fusprob}
\end{equation}

where N\textsubscript{D} and n\textsubscript{D} are the total number and number density of D atoms in the high-field zone respectively, \(\sigma_{DD}\) is the D-D fusion cross section at the center-of-mass energy corresponding to the screening energy, and \emph{x} is the linear span of the high-field zone, here assumed to be 1 nm. At a screening energy of \textasciitilde{} 8 keV we can estimate a D-D cross section of \textasciitilde{} 1 microbarn (10\textsuperscript{-30} cm\textsuperscript{2}).\textsuperscript{19,20}

Assuming a tip radius of curvature of 1 nm, a van der Waals radius of 0.18 nm for the sulfur headgroup of the dodecanethiol-d25, and close-packing of the ligand on the Au NBP tips, N\textsubscript{D} is 3086.42 D atoms loaded on each NBP, leading to a fusion probability of \ensuremath{2.47\times 10^{-10}} per pulse per NBP. Given the nominal repetition rate of 76 Mhz for the aforementioned Mai Tai® femtosecond laser system, the rate per NBP would be 0.02 s\textsuperscript{-1}. This should be observable in an experimental setting, especially considering that the \ensuremath{5\times 10^{-6}} m diameter laser spot may include anywhere from \textasciitilde10\textsuperscript{2} to 10\textsuperscript{5} Au NBPs for colloidal solutions or solid superlattices respectively.

\begin{figure}

{\centering \includegraphics[width=0.99\linewidth]{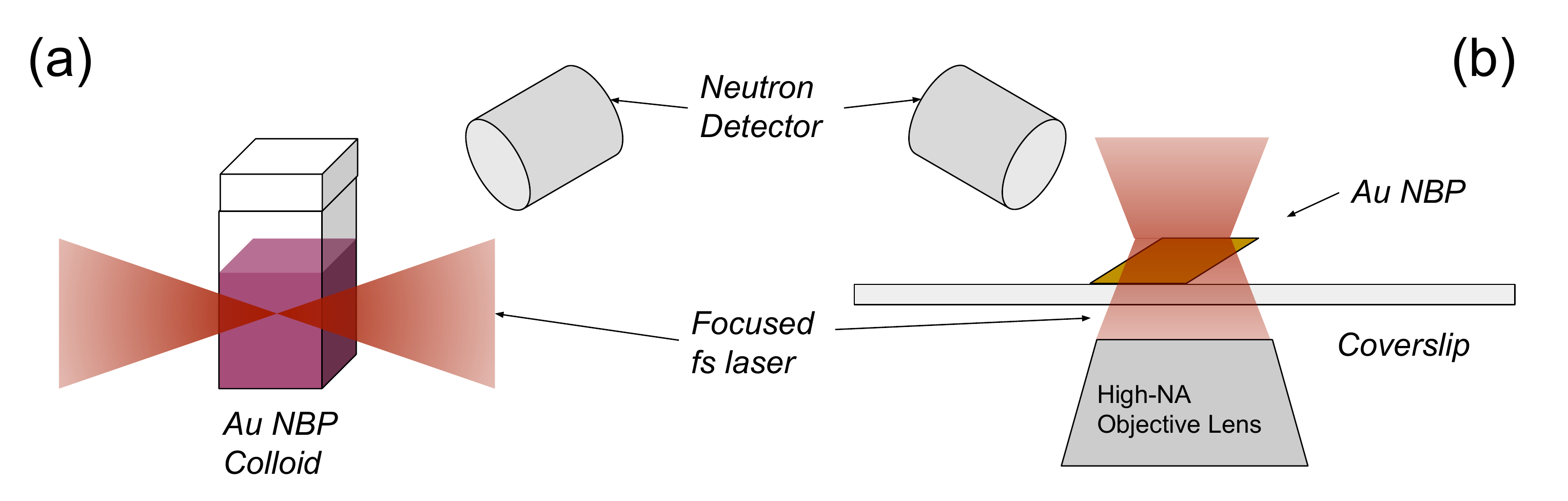} 

}

\caption{Proposed experimental scheme. (a) Excitation of bulk Au NBP colloid. (b) Excitation of single Au NBP in inverted microscope.}\label{fig:ExpFigure}
\end{figure}

A few possible experimental implementations are shown in Figure \ref{fig:ExpFigure}. In Figure \ref{fig:ExpFigure}(a), the femtosecond laser source is focused using a lens or similar optic into a cuvette containing a colloidal solution of Au NBP's. Focusing with conventional lenses will lead to a spot size of \textasciitilde10's of microns, which is consistent with the incident optical intensity used in the previous calculation. Given both the low background population of free neutrons and their long range, detectors could be placed outside the experiment to detect neutrons produced by D-D fusion. The cuvette containing the Au NP's could be surrounded with polyethylene or similar materials to thermalize any fast neutrons produced. In addition, background radiation produced by the activation of \textsuperscript{197}Au to \textsuperscript{198}Au and subsequent decay could be used as a secondary confirmation.

Alternatively as shown in Figure \ref{fig:ExpFigure}(b), an inverted microscope with a high-NA objective lens could be used to focus the incident light onto a diffraction-limited spot size (\textasciitilde800 nm), which would significantly increase the intensity. With the appropriate characterization methods such as in-situ atomic force microscopy, studies on both single and multiple closely-spaced nanoparticles could be carried out, allowing for the direct correlation of nanostructure to fusion reactivity.

\hypertarget{conclusions}{%
\section{Conclusions}\label{conclusions}}

Here we have proposed a new scheme for achieving D-D nuclear fusion in a nanoscale geometry by utilizing specially designed plasmonic metal nanomaterials (Au NBPs) and nanojoule femtosecond optical pulses from commercially available instruments. The scheme utilizes the intense electromagnetic field concentration provided by the Au NBP nanostructure and resulting ponderomotive force, which has already been shown to be capable of producing highly non-equilibrium electron densities in other plasmonic nanostructures.\textsuperscript{6--8} However, it should be noted that in most previous experiments, electric field enhancement values were well below those predicted here by at least a factor of 4. The unique sharp-tipped structure of the Au NBP, with a tip radius of curvature as low as \textasciitilde{} 1 nm, can be synthesized with high monodispersity using colloidal chemical methods. The combination of large electric field enhancement with facile lab bench synthesis separates the Au NBP from other plasmonic nanomaterials, and uniquely enables its possible application towards nuclear fusion.

While we argue that keV level screening energies will be present in this system based on the ponderomotive force, it is also important to emphasize that the nonlinear response of plasmonic nanomaterials is still an active subject of study, and other significant effects may be present to enhance the screening energy. For instance, recent work has shown that under similar conditions proposed in this work, free electrons may be ejected from the surface of the Au nanostructure, which in turn can seed the formation of a 10\textsuperscript{21} cm\textsuperscript{-3} density nano-plasma in the surrounding liquid medium.\textsuperscript{21--23} Extending this concept into a solid matrix, nano-plasmas of these densities might meet the threshold recently proposed for the formation of a micro-bubble, which upon implosion generates ultra-high electric fields and relativistic proton acceleration.\textsuperscript{24}

The electric field enhancement values shown here can be multiplied yet more by combining two or more Au NBP's tip-to-tip, similar to the ``nanobowtie antenna'' configuration.\textsuperscript{25,26} Micron- and even millimeter-scale superlattice arrays of plasmonic colloidal nanomaterials may be fabricated using various self-assembly techniques.\textsuperscript{27,28} Mixed superlattices with multiple types of nanoparticles ``dopants'' may add additional functionality. For example, Pd nanoparticles could be used to supply a continuous stream of D\textsubscript{2} fusion fuel to the Au NBP's, while various scintillator materials could convert charged particles emitted by the D-D reaction (protons, tritons and \textsuperscript{3}He) into optical excitation, thereby providing positive feedback to the incident optical field.

Finally, combining the aforementioned nanostructural engineering with nonlinear forms of optical excitation that include chirp, circular polarization\textsuperscript{29} and/or orbital angular momentum\textsuperscript{30} will allow for fine control over the timing, phase, and spatial localization of electromagnetic energy within the optical pulse envelope, thus allowing for fine control over nanometer-scale optical excitations at fusion-relevant screening energies.

\begin{acknowledgement}

The authors thank Prof. Mark Capelli, Dr. Ted Cremer and Dr. Chunching Shih for helpful discussions and comments on previous versions of this manuscript. The authors further acknowledge the use of instruments at the Electron Imaging Center for NanoMachines supported by NIH (1S10RR23057 to ZHZ) and CNSI at UCLA.

\end{acknowledgement}

\hypertarget{references}{%
\subsection*{References}\label{references}}
\addcontentsline{toc}{subsection}{References}

\hypertarget{refs}{}
\leavevmode\hypertarget{ref-enoch2012plasmonics}{}%
(1) Enoch, S.; Bonod, N. \emph{Plasmonics: From basics to advanced topics}; Springer, 2012; Vol. 167.

\leavevmode\hypertarget{ref-raether2013surface}{}%
(2) Raether, H. \emph{Surface plasmons on smooth and rough surfaces and on gratings}; Springer-Verlag Berlin An, 2013.

\leavevmode\hypertarget{ref-liu2007optical}{}%
(3) Liu, M.; Guyot-Sionnest, P.; Lee, T.-W.; Gray, S. K. \emph{Physical Review B} \textbf{2007}, \emph{76} (23), 235428.

\leavevmode\hypertarget{ref-liau2001ultrafast}{}%
(4) Liau, Y.-H.; Unterreiner, A. N.; Chang, Q.; Scherer, N. F. \emph{The Journal of Physical Chemistry B} \textbf{2001}, \emph{105} (11), 2135--2142.

\leavevmode\hypertarget{ref-jin2005correlating}{}%
(5) Jin, R.; Jureller, J. E.; Kim, H. Y.; Scherer, N. F. \emph{Journal of the American Chemical Society} \textbf{2005}, \emph{127} (36), 12482--12483.

\leavevmode\hypertarget{ref-irvine2006surface}{}%
(6) Irvine, S.; Elezzabi, A. \emph{Physical Review A} \textbf{2006}, \emph{73} (1), 013815.

\leavevmode\hypertarget{ref-irvine2005ponderomotive}{}%
(7) Irvine, S.; Elezzabi, A. \emph{Applied Physics Letters} \textbf{2005}, \emph{86} (26), 264102.

\leavevmode\hypertarget{ref-dombi2013ultrafast}{}%
(8) Dombi, P.; Hörl, A.; Rácz, P.; Márton, I.; Trügler, A.; Krenn, J. R.; Hohenester, U. \emph{Nano letters} \textbf{2013}, \emph{13} (2), 674--678.

\leavevmode\hypertarget{ref-raiola2002electron}{}%
(9) Raiola, F.; Migliardi, P.; Gang, L.; Bonomo, C.; Gyürky, G.; Bonetti, R.; Broggini, C.; Christensen, N.; Corvisiero, P.; Cruz, J.; others. \emph{Physics Letters B} \textbf{2002}, \emph{547} (3-4), 193--199.

\leavevmode\hypertarget{ref-czerski2006experimental}{}%
(10) Czerski, K.; Huke, A.; Heide, P.; Ruprecht, G. In \emph{The 2nd international conference on nuclear physics in astrophysics}; Springer, 2006; pp 83--88.

\leavevmode\hypertarget{ref-wong2019submicron}{}%
(11) Wong, A. Y. Submicron fusion devices, methods and systems. US Patent 10453575, 2019.

\leavevmode\hypertarget{ref-wong2019approach}{}%
(12) Wong, A. Y.; Shih, C.-C. \emph{arXiv preprint arXiv:1908.11068} \textbf{2019}.

\leavevmode\hypertarget{ref-liu2005mechanism}{}%
(13) Liu, M.; Guyot-Sionnest, P. \emph{The Journal of Physical Chemistry B} \textbf{2005}, \emph{109} (47), 22192--22200.

\leavevmode\hypertarget{ref-fang2016facile}{}%
(14) Fang, C.; Zhao, G.; Xiao, Y.; Zhao, J.; Zhang, Z.; Geng, B. \emph{Scientific reports} \textbf{2016}, \emph{6}, 36706.

\leavevmode\hypertarget{ref-draine1994discrete}{}%
(15) Draine, B. T.; Flatau, P. J. \emph{JOSA A} \textbf{1994}, \emph{11} (4), 1491--1499.

\leavevmode\hypertarget{ref-nanoDDSCAT}{}%
(16) Jain, P. K.; Sobh, N.; Smith, J.; Sobh, A. N.; White, S.; Faucheaux, J.; Feser, J. NanoDDSCAT, 2014.

\leavevmode\hypertarget{ref-DDSCATconvert}{}%
(17) Feser, J.; Sobh, A. N. DDSCAT convert: A target generation tool, 2013.

\leavevmode\hypertarget{ref-kaufman2007probing}{}%
(18) Kaufman, E. D.; Belyea, J.; Johnson, M. C.; Nicholson, Z. M.; Ricks, J. L.; Shah, P. K.; Bayless, M.; Pettersson, T.; Feldotö, Z.; Blomberg, E.; others. \emph{Langmuir} \textbf{2007}, \emph{23} (11), 6053--6062.

\leavevmode\hypertarget{ref-vonEngel1961}{}%
(19) Von-Engel, A.; Goodyear, C. C. \emph{Proc. Royal Society (London), Series A} \textbf{1961}, \emph{264} (1319), 445.

\leavevmode\hypertarget{ref-li2004fusion}{}%
(20) Li, X. Z.; Liu, B.; Chen, S.; Wei, Q. M.; Hora, H. \emph{Laser and Particle Beams} \textbf{2004}, \emph{22} (4), 469--477.

\leavevmode\hypertarget{ref-boulais2012plasma}{}%
(21) Boulais, E.; Lachaine, R.; Meunier, M. \emph{Nano letters} \textbf{2012}, \emph{12} (9), 4763--4769.

\leavevmode\hypertarget{ref-boulais2013plasma}{}%
(22) Boulais, E.; Lachaine, R.; Meunier, M. \emph{The Journal of Physical Chemistry C} \textbf{2013}, \emph{117} (18), 9386--9396.

\leavevmode\hypertarget{ref-labouret2016nonthermal}{}%
(23) Labouret, T.; Palpant, B. \emph{Physical Review B} \textbf{2016}, \emph{94} (24), 245426.

\leavevmode\hypertarget{ref-murakami2018generation}{}%
(24) Murakami, M.; Arefiev, A.; Zosa, M. \emph{Scientific reports} \textbf{2018}, \emph{8} (1), 1--10.

\leavevmode\hypertarget{ref-nome2009plasmonic}{}%
(25) Nome, R. A.; Guffey, M. J.; Scherer, N. F.; Gray, S. K. \emph{The Journal of Physical Chemistry A} \textbf{2009}, \emph{113} (16), 4408--4415.

\leavevmode\hypertarget{ref-malachosky2014gold}{}%
(26) Malachosky, E. W.; Guyot-Sionnest, P. \emph{The Journal of Physical Chemistry C} \textbf{2014}, \emph{118} (12), 6405--6412.

\leavevmode\hypertarget{ref-shevchenko2006structural}{}%
(27) Shevchenko, E. V.; Talapin, D. V.; Kotov, N. A.; O'Brien, S.; Murray, C. B. \emph{Nature} \textbf{2006}, \emph{439} (7072), 55--59.

\leavevmode\hypertarget{ref-shi2016two}{}%
(28) Shi, Q.; Si, K. J.; Sikdar, D.; Yap, L. W.; Premaratne, M.; Cheng, W. \emph{ACS nano} \textbf{2016}, \emph{10} (1), 967--976.

\leavevmode\hypertarget{ref-hentschel2017chiral}{}%
(29) Hentschel, M.; Schäferling, M.; Duan, X.; Giessen, H.; Liu, N. \emph{Science advances} \textbf{2017}, \emph{3} (5), e1602735.

\leavevmode\hypertarget{ref-sakai2018nanofocusing}{}%
(30) Sakai, K.; Yamamoto, T.; Sasaki, K. \emph{Scientific reports} \textbf{2018}, \emph{8} (1), 7746.
\end{document}